\documentclass[10pt, conference, letterpaper]{ IEEEtran}
\IEEEoverridecommandlockouts

\newif\ifcomment
\commenttrue

\ifcomment
    \newcounter{ALNumberOfComments}
    \stepcounter{ALNumberOfComments}
    \newcounter{GINumberOfComments}
    \stepcounter{GINumberOfComments}

    \newcommand{\NOTE}[1]
    {
      {\footnotesize\it
        \begin{center}
          \begin{tabular}{|c|}
           \hline
            \parbox{0.85\columnwidth}{
              \medskip
              #1
              \medskip} \\
            \hline
          \end{tabular}
        \end{center}
        }
    }
\else
    \newcommand\NOTE[1]{}
\fi

\pdfoutput=1

\def\sumT{\sum_{t=1}^T}
\def\E{\mathbb E}
\newcommand{\dtp}[2]{\langle {#1}, {#2} \rangle}

\usepackage{cite}
\usepackage{xcolor}
\usepackage{graphicx}
\usepackage{amssymb}
\usepackage{amsmath}
\usepackage{epstopdf}
\usepackage{mathtools}
\usepackage{url}
\usepackage{wrapfig}
\usepackage{caption}
\usepackage{enumitem}
\usepackage{mdframed}
\usepackage{tcolorbox}
\usepackage{subcaption}
\usepackage{bm}
\usepackage[ruled,commentsnumbered]{algorithm2e}
\usepackage{amsfonts}
\usepackage{dsfont}
\usepackage{lipsum}
\usepackage{booktabs}  

\renewcommand{\baselinestretch}{0.96}

\def\BibTeX{{\rm B\kern-.05em{\sc i\kern-.025em b}\kern-.08em
    T\kern-.1667em\lower.7ex\hbox{E}\kern-.125emX}}

\title{Smooth Handovers via Smoothed Online Learning}

\usepackage{authblk}
\author{Michail Kalntis\textsuperscript{*}, Andra Lutu\textsuperscript{\textdagger}, Jes\'us Oma\~na Iglesias\textsuperscript{\textdagger}, Fernando A. Kuipers\textsuperscript{*}, George Iosifidis\textsuperscript{*}}

\affil{\textsuperscript{*}Delft University of Technology, The Netherlands. Emails: \{m.kalntis, f.a.kuipers, g.iosifidis\}@tudelft.nl}

\affil{\textsuperscript{†}Telef\'onica Research, Spain. Emails: \{andra.lutu, jesusalberto.omana\}@telefonica.com}

\begin{document}
\maketitle
\begin{abstract}
With users demanding seamless connectivity, handovers (HOs) have become a fundamental element of cellular networks. However, optimizing HOs is a challenging problem, further exacerbated by the growing complexity of mobile networks. 
This paper presents the first countrywide study of HO optimization, through the prism of Smoothed Online Learning (SOL).
We first analyze an extensive dataset from a commercial mobile network operator (MNO) in Europe with more than 40M users, to understand and reveal 
important features and performance impacts on HOs. Our findings highlight a correlation between HO failures/delays, and the characteristics of radio cells and end-user devices, showcasing the impact of heterogeneity in mobile networks nowadays. 
We subsequently model UE--cell associations as dynamic decisions and propose a realistic system model for smooth and accurate HOs that extends existing approaches by (i) incorporating device and cell features on HO optimization, and (ii) eliminating (prior) strong assumptions about requiring future signal measurements and knowledge of end-user mobility. Our algorithm, aligned with the O-RAN paradigm, provides robust dynamic regret guarantees, even in challenging environments, and shows superior performance in multiple scenarios with real-world and synthetic data.
\end{abstract}

\section{Introduction}

\textbf{Motivation}. Mobility management has been a prime consideration for every generation of mobile networks, and occupies a conspicuous position in the agendas of industry and academia \cite{ho-tutorial-2019, ho-optimization-survey}. However, as the proposed solutions evolve, so does the problem. On the one hand, users and their devices (user equipments, i.e., UEs) are becoming increasingly mobile and expect higher-data rates and lower-latency services. On the other hand, today's networks are more heterogeneous than ever, encompassing different Radio Access Technology (RAT) generations,  multi-tier cells (some with small coverage) / base stations (BSs) operating a wide range of frequencies, and equipment from multiple vendors. This heterogeneity and complexity compound mobility management issues and, in particular, perplex the UE-BS association decisions and the induced handovers (HOs).

HOs are fundamental elements for enabling seamless connectivity for mobile users. Optimizing this procedure is crucial, as inefficient or suboptimal HOs can have a dire twofold impact: on the network-side, they might lead to increased resource consumption and signaling overhead \cite{andrews-towards}; and on the UE-side, they may cause service interruptions and battery depletion \cite{vivisecting_2022, pp_ho_2023}.
Thus, latest proposals depart from conventional signal-to-noise ratio (SNR) or signal-to-interference-plus-noise ratio (SINR) based association rules, towards solutions where UE-BS associations are based on network-wide criteria, e.g., aggregate throughput or fairness/load-balancing metrics, while also aiming to reduce HOs whenever possible \cite{andrews-association, andrews-globecom21, RL-TWC22,  kelleler-jsac23}. The recent Open Radio Access Network (O-RAN) paradigm \cite{oran-andres} facilitates such holistic approaches and enables central \emph{controllers} to coordinate the network operations dynamically \cite{o-ran-hos-tmc24, tassiulas-globecom}, including how the UE associations are implemented.

In this context, the goal of this paper is to design a novel UE association and smooth HO control mechanism that is both \emph{effective} and \emph{robust}, thus overcoming key limitations of the above works (see discussion in Sec. \ref{sec:related-work}). The effectiveness alludes to achieving a network utility maximization criterion with guarantees under a wide range of conditions, despite the heterogeneity of UEs and BSs. The robustness, on the other hand, refers to ensuring this performance without the controller requiring access to accurate information about the UEs/network-related parameters, such as their mobility patterns. These two main features of our proposal fill a key gap in the literature and pave the road for the next generation of mobility management solutions.

\begin{table}[!t]
	{
		\small{
			\centering
			\caption{Mobility Dataset Size for European Country.}
			\vspace{-2mm}
			\begin{tabular}{p{3.98cm}p{3.6cm}}
				
				\toprule
				\# of cell sites, \# of base stations & $ > 26$\small{k},  $> 370$\small{k} \\
				\# of UEs measured & $\approx$ 40\small{M} \\
				\# handovers (daily) & $ > 1.7$\small{B}\\
				Measurement duration & 7 days ($\approx$ 8 TB/day) \\
    \bottomrule
			\end{tabular}
			\label{tab:statistics}
			\vspace{-6mm}
		}
	}
\end{table}

\noindent\textbf{Methods \& Contributions}. To understand HOs and their effects in practice, we kick off our study with the analysis of a 1-week capture of all mobility events from a large mobile network operator (MNO) in a European country (see Table \ref{tab:statistics}). The analysis of this vast dataset sheds light on the frequency, features, and performance of HOs. For instance, we find that HO failures and delays depend heavily on the RAT generation, and are affected by the UE type. In other words, \emph{HO events vary widely in terms of completion delay}. We also reveal the network complexity in terms of RATs and load in this top-tier operator and discuss the implications of this heterogeneity in terms of HOs. 
Finally, we measure the effect of HOs on several network and user key performance indicators (KPIs), such as packet loss and throughput. To the best of our knowledge, this is the first \emph{network-side HO optimization and measurement} study that overcomes the limitations of prior UE-side studies, or large network-side measurement-only HO study \cite{kalntis_imc24}. 

Building on these findings, we introduce a realistic system model that captures the impact of the different HOs and mitigates them while maximizing the network throughput. Our model aligns with recent works~\cite{andrews-association,tassiulas-globecom, andrews-globecom21,kelleler-jsac23,choi-TWC15, o-ran-hos-tmc24}, which we extend substantially by accounting for the network and HO diversity, and importantly, by dropping requirements for access to future SINRs and UE mobility patterns. It is commonly accepted that this is a strict condition that limits the applicability of such solutions. Additionally, we do not assume that the relevant UE/BS parameters are stationary-perturbed since rapid and unpredictable channel fluctuations are becoming increasingly common in heterogeneous networks and mobile services. In fact, our perturbation model is an adversarial one, where the various random parameters can even be selected by an attacker; still, all results and guarantees hold.

With this in mind, we turn the problem on its head and study HOs through the lens of online convex optimization (OCO)~ \cite{hazan-book}. Namely, we model the UE-BS associations as dynamic decisions that the network controller updates in a time-slotted fashion, where successive (de)associations induce undesirable (sometimes necessary) HOs. We argue a natural framework for this setting is that of \emph{smoothed online learning} (SOL) \cite{wierman-smoothed2018}, which maximizes a performance criterion while reducing the decision changes. Using such a framework enables the controller to be oblivious to the SINRs for each UE-BS pair when deciding the associations, and the throughput these will achieve; indeed, it is challenging to predict accurately or know the SINRs over a time window of several msec/sec  \cite{kalntis_tcom24}. Still, following a rigorous analysis, we show that our learning algorithm ensures \emph{sublinear dynamic regret}, i.e., its gap w.r.t. an ideal oracle that has full information about the future diminishes with time \cite{zhang-smoothed-ol}. Our model is informed by, and aligned with, the O-RAN paradigm, and the proposed algorithm can be implemented as xApp in \textit{near-real-time} (running time 10msec--1sec) \cite{oran-andres}; the reader is kindly referred to \cite{kalntis22, kalntis_tcom24} for the involved interfaces and detailed steps. Finally, to verify the robustness of our proposed solution, we evaluate its performance on simple and extreme (i.e., adversarial) synthetic scenarios in accordance with related work, as well as in real scenarios using signal measurement from crowdsourced data (i.e., measured on the field). 

In summary, the contributions of this work are:

\noindent$\bullet$ We present HO and network statistics by using a 1-week dataset from a tier-1 MNO. We highlight the network heterogeneity and identify key factors impacting smooth HOs. We are the first to optimize HOs from the MNO's perspective, which sets the basis for realistic HO optimization models.

\noindent$\bullet$ We model the HO optimization as a \emph{smoothed online learning problem} where the HO delays depend on the RAT and UE type, and assume no prior information for the channels and mobility patterns. This approach departs from related work and its (often simplifying) modeling assumptions.

\noindent$\bullet$ We design a scalable (near-real-time) algorithm that achieves sublinear dynamic regret, and we characterize its performance w.r.t. system parameters. We also propose extensions for the case of time-varying HO delays and the case of available (untrusted) forecasting tools. 

\noindent$\bullet$ We create a simulator using our real, crowdsourced radio signal quality measurements, as well as actual cell and UE information, and evaluate our solution against meaningful benchmarks; e.g., we find up to $\times79.6$ lower HO cost than previous works, without sacrificing throughput.
\vspace{-0.5mm}
\section{Background \& Related Work} \label{sec:related-work}

\noindent\textbf{Measurements}. HOs have mainly been studied using traces from UEs, which are inevitably limited to certain manufacturers \cite{vivisecting_2022}, services \cite{yuan2022understanding}, or user types  \cite{wheels_2023, polycorn_trains23}. Our findings on HO duration align with these prior works \cite{wheels_2023, vivisecting_2022, lte_rails_2019} and offer new insights into the impact of RAT and UE type on HOs. While several studies have measured HO volume \cite{yuan2022understanding, 5G_ues_2020, 5G_wild_2021}, noting that horizontal HOs are more frequent in 5G-Standalone (5G-SA) and 4G, and vertical HOs in 5G-NSA \cite{yuan2022understanding}, we enhance these findings by revealing the geographic heterogeneity of HOs and measuring their effect on important KPIs. Importantly, our work utilizes a large-scale dataset measuring HOs from the network-side of a top-tier operator with 40M UEs. This contrasts sharply with the few network-side studies, e.g., \cite{pp_ho_2023} has less than 0.7\% of our UEs. While \cite{kalntis_imc24} is the first countrywide study of cellular HOs, it focuses solely on measurements, without providing a HO solution or focusing on the effect of HO failures (HOFs) on KPIs.

\noindent\textbf{Handover Optimization}. In user-centric mobility schemes, UE associations (and thus HOs) are often approached as a bandit learning problem where each UE explores which BS offers better throughput \cite{schaar-jsac16, schaar-wcnc17, schaar-TWC19, shroff-iwqos21, tao-TMC21, kasbekar-TVT22}. Conversely, using SINR measurements from all UEs, the network can make more effective centralized decisions. This idea is regaining momentum, but is also more challenging (due to scale, among others) to implement. For instance, a recent thread of studies employs (Deep) Reinforcement Learning to decide HOs or HO rules (e.g., SINR thresholds) \cite{iot-2018, RL-TNSM-20,RL-TWC22,andrews-globecom21, kelleler-jsac23, o-ran-hos-tmc24}. Despite its modeling appeal, these approaches essentially rely on heuristics. On the other hand, our method comes with performance guarantees, even under adversarial conditions.

Similarly, the joint optimization of throughput and HO delays is considered in \cite{andrews-association, andrews-globecom21, kelleler-jsac23, choi-TWC15, o-ran-hos-tmc24}. Our goal is very close to these works, but we enrich their model to capture the different delay effects of the various HO types, RATs, and UE types. Also, we drop several of their assumptions and, in particular, their need to have accurate SINR information. Clearly, in today's volatile and non-stationary network conditions, this assumption is impractical. Finally, some recent papers leverage the ML-provisions of O-RAN, and propose forecasting-assisted HOs \cite{tassiulas-globecom, masri-ml-ho,andrews-association, o-ran-hos-tmc24}. Our solution is also aligned with O-RAN, but unlike these works, it learns on the fly the network/UE parameters and does not require offline training nor the availability of a reliable forecaster. At the same time, when such predictions are available, our algorithm can directly benefit from them to expedite its learning, building on the idea of optimistic learning \cite{rakhlin-optimistic}. For additional discussion on HOs, see \cite{ho-optimization-survey, ho-tutorial-2019}.

\noindent\textbf{Smoothed Online Learning}. In terms of solution, unlike the model-predictive control of \cite{andrews-association}, the RL strategies in \cite{andrews-globecom21, kelleler-jsac23, zorzi-twc23, o-ran-hos-tmc24}, or the heuristics in \cite{choi-TWC15}, we approach this problem, for the first time, through the lens of \emph{smoothed online learning} \cite{wierman-smoothed2018, zhang-18, zhang-smoothed-ol}. SOL enriches the online convex optimization toolbox \cite{hazan-book}, accounting for costs induced by decision changes, and has recently found applications in caching \cite{abishek-caching, ton-caching}, network selection \cite{schaar-TWC19}, and service deployment \cite{steiger-infocom23}, among others. Here, the decisions express the UE-BS associations, and the switching cost captures the HO delays in a natural way. Yet, unlike the above works, we leverage the more competitive dynamic regret benchmark, since HOs are unavoidable, and the switching cost models the delay rather than the mere HO count. We also note that SOL should not be confused with mathematical smoothing \cite{rakhlin-smoothed-colt22}, nor with the stochastic and full-information framework of decisions with reconfiguration delays \cite{eytan-reconfigToN15}.

\section{Data collection \& analysis} \label{sec:data-analysis}

\noindent\textbf{Data collection}. To show the network's heterogeneity and pinpoint key factors affecting seamless HOs (and thus, are essential for our system model), we collect passive measurements using commercial tools available within the MNO's infrastructure. For that, we focus on elements in different RATs, such as the Mobile Management Entity (MME), Mobile Switching Center (MSC), Serving GPRS Support Node (SGSN), Serving Gateway (SGW), and cell sites.
The captured data spans the entire country from 24-June-2024 to 30-June-2024 (Table \ref{tab:statistics}) and includes the \emph{(i)} HO timestamp, \emph{(ii)} HO duration (msec granularity), \emph{(iii)} anonymized user ID, based on the International Mobile Subscriber/Equipment Identity (IMSI/IMEI), and \emph{(iv)} source and target BS of the HO along with their respective RATs. 
At the time this study was conducted, the majority of the traffic for the studied MNO used the 5G-Non-Standalone (NSA) deployment \cite{SA_vs_NSA}.
From a mobility management perspective, 5G-NSA and 4G are identical, since the former relies on the 4G Evolved Packet Core (EPC); thus, we use the term ``4G/5G-NSA''. 

In addition, we leverage \emph{(i)} a daily-captured dataset from the MNO that contains cell information such as the longitude, latitude, vendor, and supported RATs, \emph{(ii)} a daily commercial device catalog, provided by the Global System for Mobile Communications (GSM) Association (GSMA) to examine correlations of user-specific characteristics with HOs, \emph{(iii)} open datasets published by the official census office in the studied European country to explore the distribution of HOs across the different geographic areas (mainly the 300+ districts defined by the census office), and \emph{(iv)} commercial crowdsourced data with msec granularity of radio signal measurements to assess the performance of our algorithm.

\begin{figure}[t]
    \centering
    \includegraphics[width=0.83\linewidth]{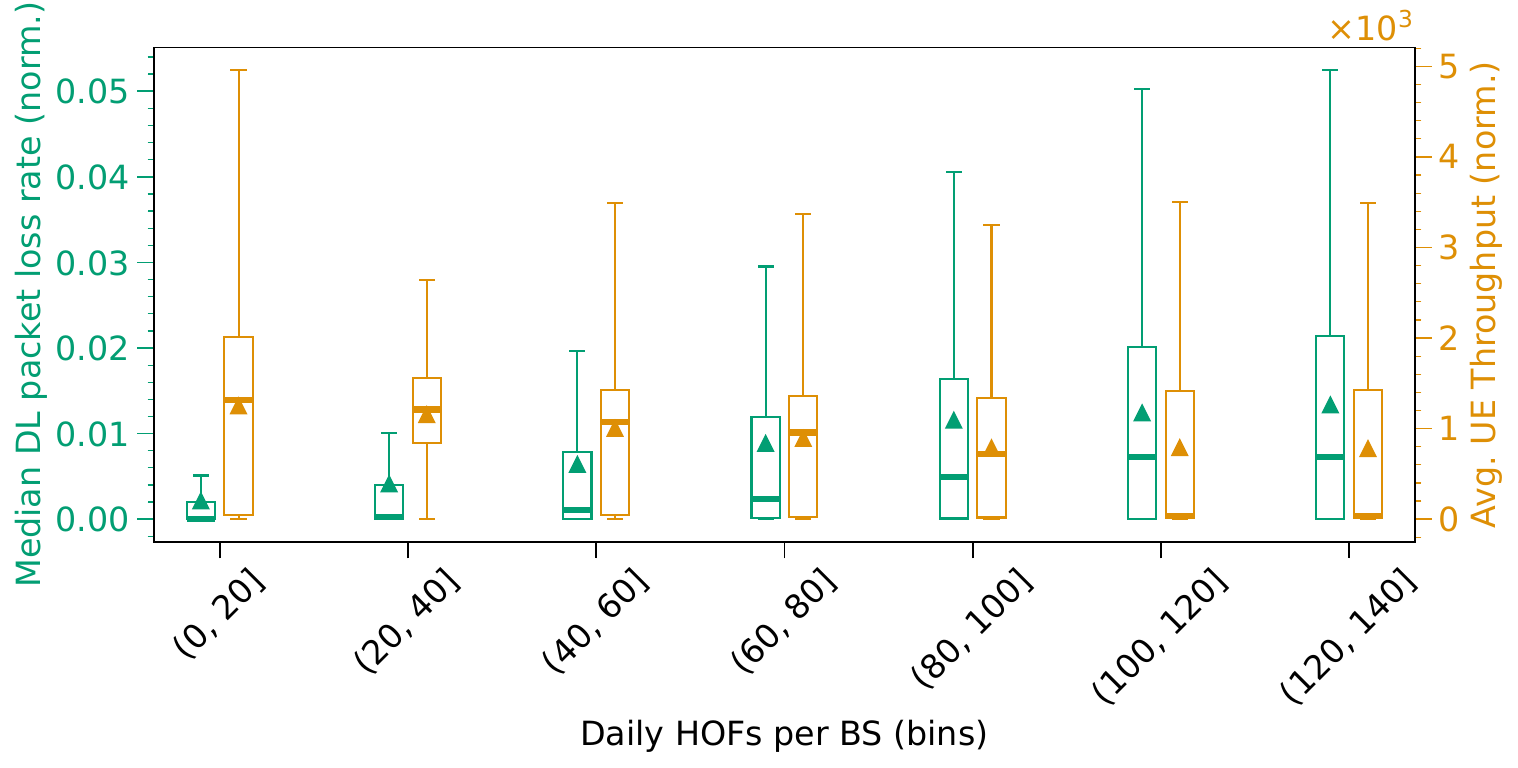}
    \vspace{-2mm}
    \caption{\small{Median normalized DL packet loss (left, green y-axis) and normalized average UE throughput (right, orange y-axis) vs binned daily HOFs. Triangles and bold horizontal lines show the mean and median, respectively, in each boxplot.}}
    \label{fig:hof_vs_kpi}
    \vspace{-6mm}
\end{figure}

\noindent\textbf{Effect of HOs on KPIs}. We first study how HOs and HOFs affect some important network KPIs, used by the operational team to determine the health of the network and reason about the quality of the service to the end-users.
For confidentiality, we normalize each KPI by its median from all days. Also, we discard the outliers, i.e., cells with daily HOFs $>$140 (0.1\% of data). Fig. \ref{fig:hof_vs_kpi} depicts the daily HOFs per cell and their impact on downlink (DL) packet loss and user throughput per day. We observe a decrease in normalized average user throughput with increasing HOFs; e.g., (0, 20] HOFs lead to a normalized mean throughput of 1.25k (normalized value, i.e., does not mean Kbps or Mbps), which declines by 37.6\% to 0.78k for (120, 140] HOFs per day. Similar studies (e.g., \cite{5G_ues_2020, wheels_2023, 5G_wild_2021}) have reported UE throughput but only for specific UE types (smartphones) and manufacturers. Simultaneously, the normalized median DL packet loss rises from a mean of around 0.002 to 0.013 when daily HOFs increase from (0, 20] to (120, 140] respectively. The higher mean compared to the median in each boxplot indicates a few UEs with larger losses compared to the majority.

We verified the findings of Fig. \ref{fig:hof_vs_kpi} with a generalized linear regression model using as dependent variable the DL packet loss rate and as independent variables the number of HOs and HOFs while controlling for all key covariates; namely, the HO type (inter/intra RAT), district population, BS type/vendor/transmission power, and the district. This ensures the packet loss effects are indeed due to HOs and HOFs, and not to some latent factor. We find that 1\% increase of HOs in a cell, increases by $0.02\%$ the packet loss rate, on average, for the served users; and even worse, 1\% increase in HOFs increases by $0.6\%$ this loss. A similar model found that, all else being equal (including total uplink/DL volume and physical resource blocks), a HOF increase of 1\% reduces by $0.008\%$ the average user throughput. And this drop is more pronounced in cells with few HOFs, as Fig. \ref{fig:hof_vs_kpi} shows.

\begin{figure}[t]
    \centering
    \includegraphics[width=0.88\linewidth]{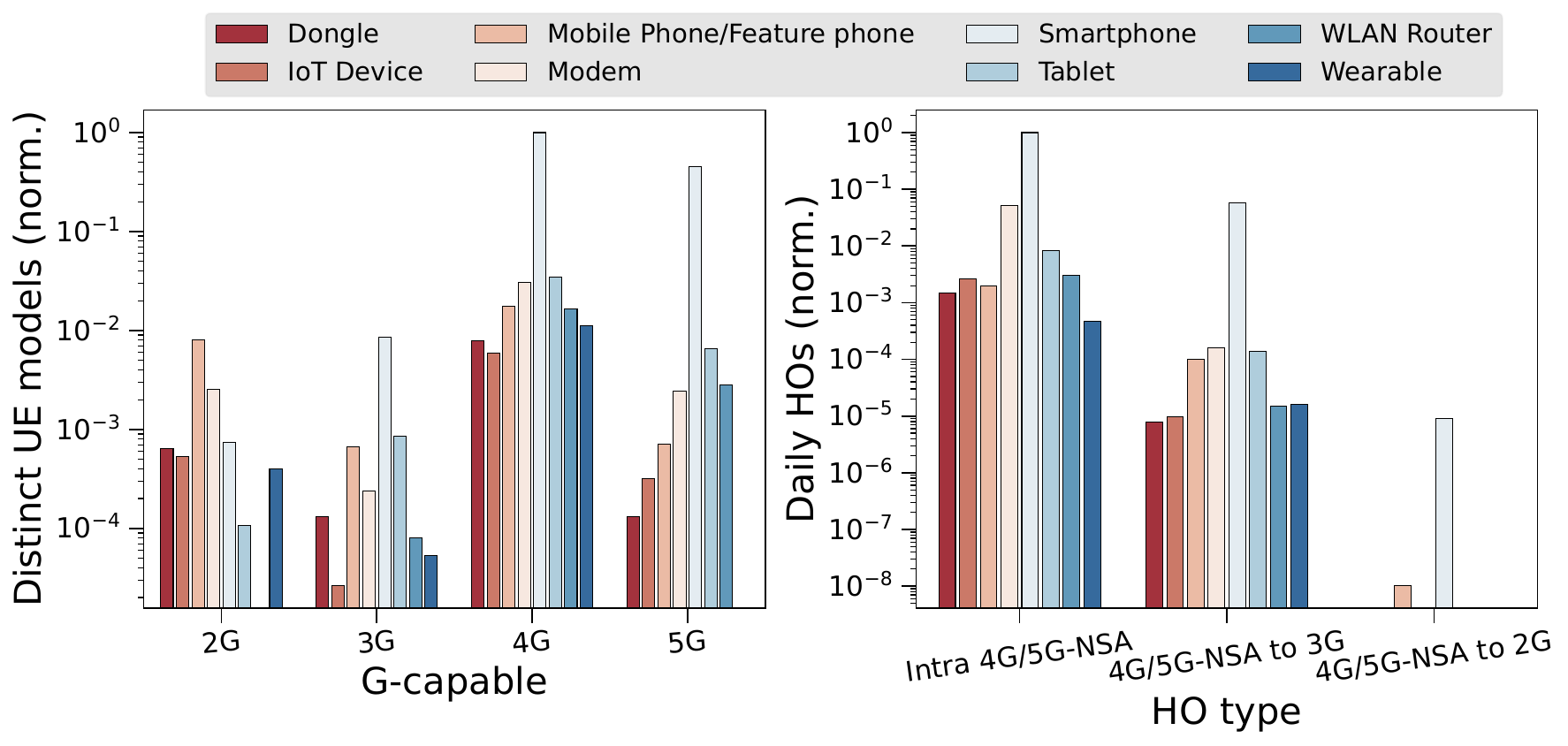}
    \vspace{-2mm}
    \caption{\small{\textbf{Left}: number (norm. by max) of different UE models and their RAT capabilities (up to 2G, 3G, 4G, 5G). \textbf{Right}: HOs (norm. by max) per day each of the UE model executes, and to what RAT.}}
    \label{fig:ue_diversity}
    \vspace{-5mm}
\end{figure}

\noindent\textbf{Network \& UE Heterogeneity}. 
Using the GSMA devices database, we discern the eight most crucial \textit{UE types} (dongle, IoT, feature phone, modem, smartphone, tablet, WLAN router and wearable), and infer their connectivity capabilities (i.e., up to what RAT they support) from their frequency bands. Given the dominance of 4G/5G capable models (98.5\%), depicted in Fig. \ref{fig:ue_diversity}, we study the HOs from 4G/5G-NSA to the same or older RATs. In these \textit{HO types} (i.e., Intra 4G/5G-NSA, 4G/5G-NSA to 3G, 4G/5G-NSA to 2G), approximately 94\% are Intra 4G/5G-NSA, caused by a wide range of devices. HOs to 3G and 2G hold an important 6\%, mainly from smartphones, modems, and tablets, magnifying the heterogeneity in these dimensions as well. As we show in the sequel, the HO delay in the older RATs is $\times5$--40 higher. Network heterogeneity is also extensively analyzed in \cite{kalntis_imc24}.

\begin{figure}[t]
    \centering
    \includegraphics[width=0.83\linewidth]{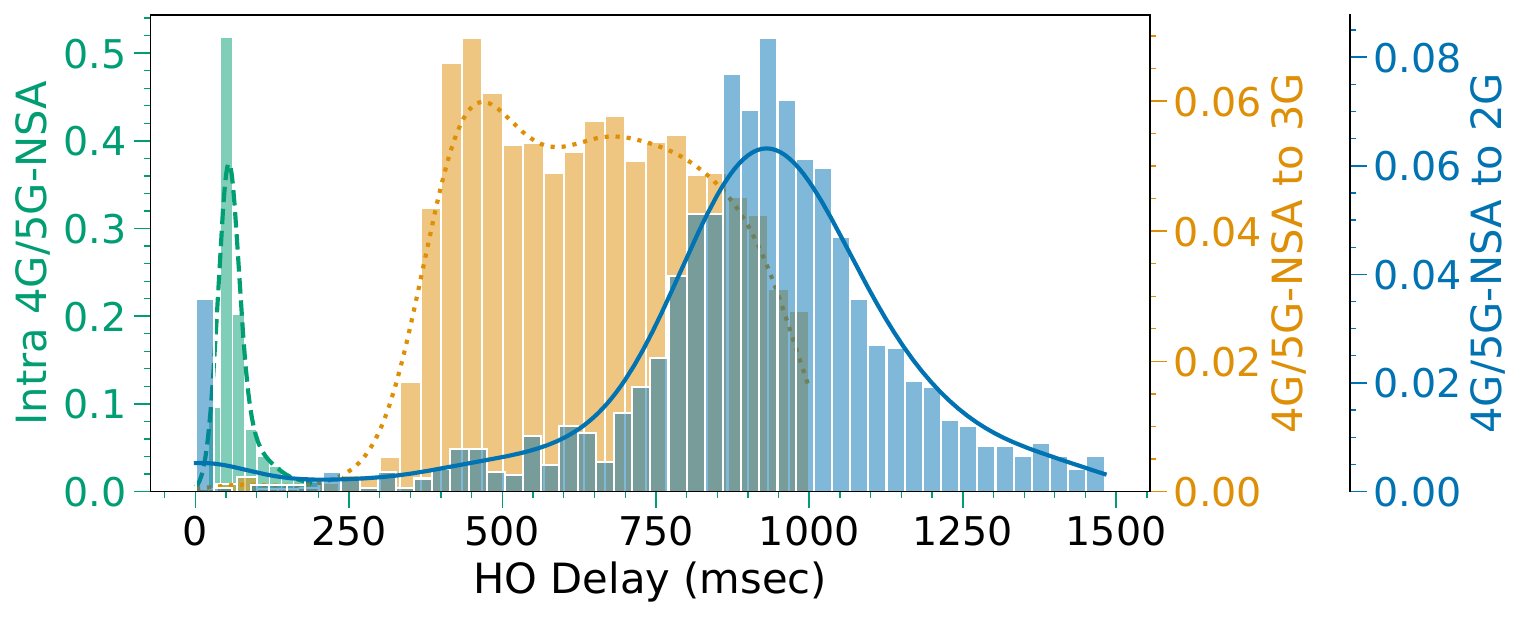}
    \vspace{-2mm}
    \caption{\small{Histogram (bars) and distribution (line) of the HO delays for each HO type (all same-colored bars sum to 1).}}
    \label{fig:RAT_ho_delays}
    \vspace{-5mm}
\end{figure}

\begin{figure}[t]
    \centering
    \includegraphics[width=0.82\linewidth]{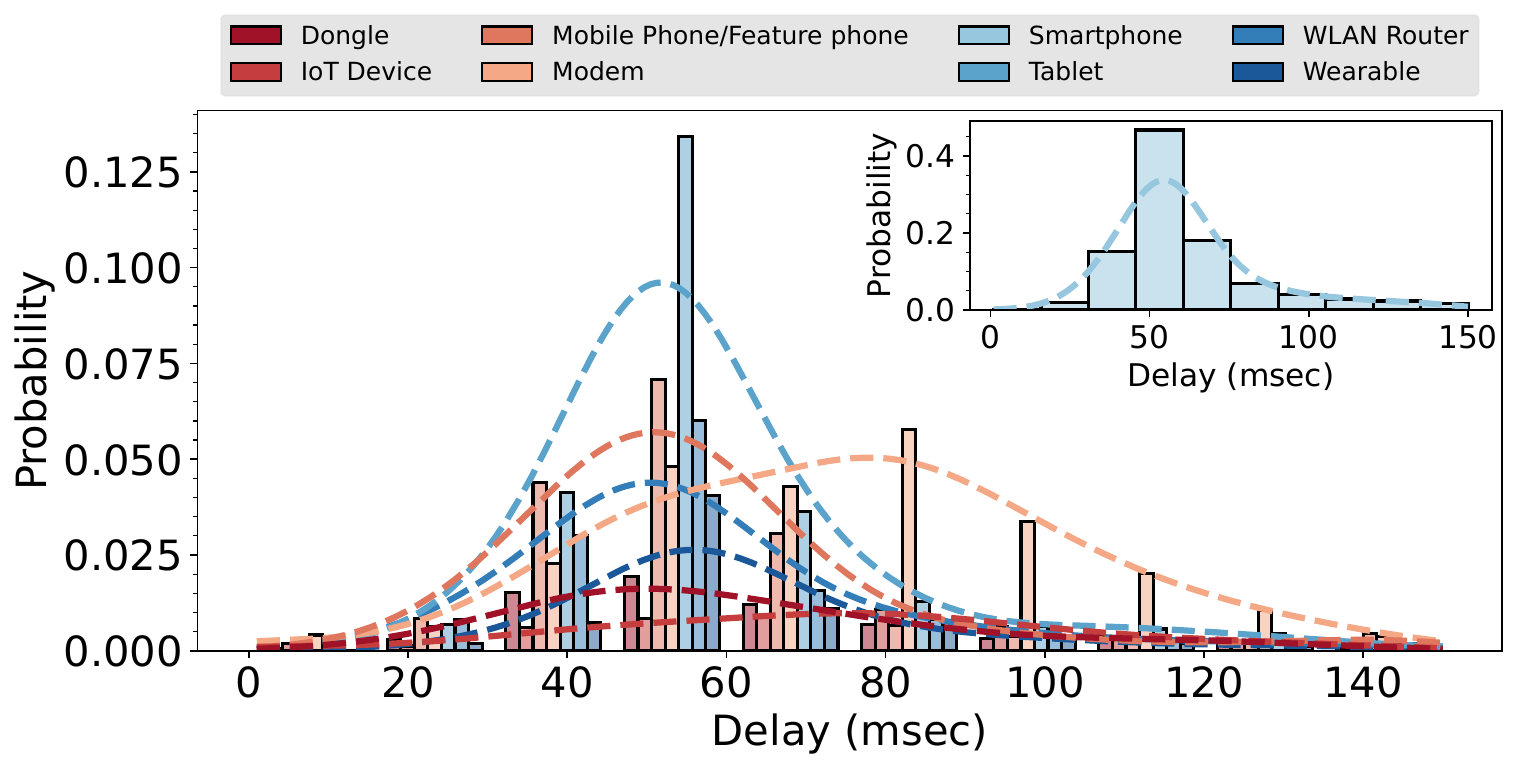}
    \vspace{-2mm}
    \caption{\small{Histogram (bars) and distribution (line) of the HO delays for each UE model (all same-colored bars sum to 1).}}
    \label{fig:ue_ho_delays}
    \vspace{-6mm}
\end{figure}

Fig. \ref{fig:RAT_ho_delays} and Fig. \ref{fig:ue_ho_delays} illustrate the histogram and probability density of different HOs and device types, respectively, w.r.t. the HO delay. From Fig. \ref{fig:RAT_ho_delays}, we observe that the median Intra 4G/5G-NSA HOs require approximately 50msec, whereas HOs to 3G range from 400 to 950msec and to 2G from 750 to 1100msec; thus, especially in the older RATs, the distribution is significantly more spread. Moreover, from Fig. \ref{fig:ue_ho_delays}, we deduct the different distributions of the HO delays for each device type. It is interesting to observe the different mean and variances of each UE type; for instance, modems and IoT devices require on average 75msec and 73msec respectively, but around the same number of these models start from 50msec and reach 110msec; smartphones, on the other hand, need 50--62 msec. Consequently, \textit{accounting both for the UE type and HO type is essential when optimizing for the HO delay}.

\section{System Model \& Problem Statement}\label{sec:model}

We consider a heterogeneous cellular network comprising a set $\mathcal J$ of $J$ base stations serving a set $\mathcal I$ of $I$ users (UEs) in the downlink. Each BS is characterized by an array of features such as its operating frequency, RAT generation (2G-5G), location, etc. Similarly, the set of UEs comprises smartphones, feature phones, IoT devices, and so on. The network is managed centrally by a network controller, in the spirit of recent O-RAN architecture proposals \cite{oran-andres}. The system operation is time-slotted where we index the time slots with $t$ and without loss of generality assume the slots have unitary length. These slots refer to the UE association intervals, which subsume other resource scheduling time-slots (e.g., for power control). We study the system for a set $\mathcal T$ of $T$ slots. The key metric for the association decisions is the SINR for the signal delivered by BS $j$ to user $i$:
\begin{align}
	s_{ij}(t)=\frac{p_j\phi_{ij}(t)}{	\sum_{k\in \mathcal{B}_j} p_k\phi_{ik}(t) + w_k\sigma^2},
\end{align}
where $p_j$ is the transmit power of BS $j$, $\mathcal B_j$ the set of BSs that operate in the same frequency as $j$, $\phi_{ij}(t)$ the channel gain (including pathloss, shadowing, and antenna gains), $w_k$ the bandwidth of BS $k$, and $\sigma^2$ the power spectral density. In line with previous works \cite{andrews-association,kelleler-jsac23, andrews-globecom21, choi-TWC15}, $s_{ij}(t)$ is the average SINR over the slot $t$ (the UEs report multiple SINRs during each slot \cite{andrews-association}).

The rate that each user $i\in \mathcal I$ associated with BS $j\in \mathcal J$ achieves during slot $t$, can be expressed as:
\begin{align}
	r_{ij}(t)\!=\!\frac{c_{ij}(t)\big(1\!-\!d_{ij}(t)\big)}{y_j(t)},  \label{eq:rate}
\end{align}
where $y_j(t)$ is the total number of UEs that BS $j$ needs to serve during $t$, and $c_{ij}(t)=w_j\log_2\big(1+ s_{ij}(t)\big)$ is the maximum possible rate for $i$ if it was using BS $j$ exclusively. The rate $r_{ij}(t)$ is discounted by the service disruption time $d_{ij}(t)$ which models any HO delay associated with the assignment of UE $i$ to BS $j$. Clearly, this delay is negligible if $i$ was already associated with BS $j$; is pronounced when the target BS is of different RAT; and is even larger when there is a HO failure (see Sec. \ref{sec:data-analysis}). We normalize these parameters, i.e., $d_{ij}(t)\!\leq\! 1, \forall i,j$, to express the \emph{portion} of slot the UE was not receiving service. 
Equation \eqref{eq:rate} assumes the BS resources are allocated fairly across the active users via, e.g., a round robin or a proportional-fair scheduler \cite{tse-scheduler}. As it will become clear, our analysis holds for different scheduling schemes.

We denote with $x_{ij}(t)\in\{0,1\}$ the association of user $i$ with BS $j$ in the beginning of slot $t$ and define the vector $\bm x_t\!=\!\big(x_{ij}(t) \!\in\! \{0,1\}, i\in\mathcal I, j\in\mathcal J\big)$. Then, the problem the network controller wishes to solve can be expressed as:
\begin{align}
 \mathbb{P}:\quad 	\max_{\{\bm{x}_t\}_{t}} \quad & \sum_{t=1}^T\sum_{i=1}^I\sum_{j=1}^Jx_{ij}(t)\log r_{ij}(t) \label{problem-obj}\\
	\textrm{s.t.} \quad & \sum_{j\in\mathcal J}x_{ij}(t)=1, \quad \ \ \forall i\in\mathcal I,  \label{problem-assignment-constr}\\
	&y_j(t)=\sum_{i\in \mathcal I} x_{ij}(t), \ \  \forall j\in \mathcal J, \label{problem-load-constr}\\
	&\bm{x}_t\in\{0,1\}^{I\cdot J}, \ \ \ \quad \ \forall t \in \mathcal T. \label{problem-binary-constr}
\end{align}
The logarithmic utility function is selected to maximize the product of rates so as to balance network sum-rate and fairness \cite{srikant-log, andrews-association, andrews-globecom21}; constraint \eqref{problem-assignment-constr} ensures each UE is assigned to one BS, and \eqref{problem-load-constr} calculates the assigned UEs to each BS.

The solution to this problem at the beginning of horizon $T$ is hindered by several factors. First, at $t=1$, the controller does not have access to all future SINR values for each UE-BS pair. In fact, the SINRs during each slot $t$ are practically unknown \emph{even at the beginning} of that slot. Secondly, the HO delays $\{d_{ij}(t)\}$ depend on the associations $\bm x_t$, but also on the previous associations $\bm x_{t-1}$, since these two vectors determine if there is a HO or not; a HO is triggered for user $i$ if $\bm x_{ij}(t-1) \neq \bm x_{ij}(t)$. In other words, there is a memory effect in the system, thus the problem cannot be decomposed on per-slot basis. With these challenges in mind, our goal is to design an online association algorithm that is oblivious to these time-varying unknown parameters, and which nevertheless maximizes the throughput while minimizing the HO delays, compared to a meaningful (i.e., competitive) benchmark.

\section{Algorithm Design and Analysis} \label{sec:algorithm}

\noindent\textbf{Reformulation \& Benchmark}. We approach $\mathbb{P}$ as a \emph{smoothed online learning} problem and tackle it via meta-learning based on the \emph{experts framework} \cite{warmuth-experts, hazan-meta}. The main idea is to deploy a set of parallel learning algorithms with different learning rates, and a meta-learner that tracks their performance and discerns on-the-fly the best-performing one. We start by reformulating the objective: 
\begin{align*}
&f_t(\bm x_t)\!\triangleq\!\sum_{i=1}^I\sum_{j=1}^J\Big[ x_{ij}(t)\log c_{ij}(t)  \!-\!  x_{ij}(t)\log y_{j}(t)\\
&\!+\!x_{ij}(t)\log \big(1\!-\!d_{ij}(t) \big)\Big] \stackrel{(\alpha)}=\!\sum_{i=1}^I\sum_{j=1}^J\! x_{ij}(t)\log c_{ij}(t)\\
& - \sum_{j=1}^J y_{j}(t)\log y_{j}(t) + \sum_{i=1}^I\sum_{j=1}^J x_{ij}(t)\log \big(1-d_{ij}(t) \big)  
\end{align*}
where $(\alpha)$ follows from $y_j(t)\!=\!\sum_{i\in\mathcal I} x_{ij}(t), \forall j, t$. The last term corresponds to the performance cost due to HO delays. Extending the rationale of prior works \cite{andrews-association, kelleler-jsac23, choi-TWC15}, and based on our data analysis, we will capture this cost using:
\begin{align}
\!	h\big(\bm{x}_t, \bm{x}_{t-1}\big)\!=\!-\gamma\left\|\bm A \big(\bm{x}_t-\bm{x}_{t-1}\big)\right\| \!\triangleq  \!- \gamma\|\bm{x}_t-\bm{x}_{t-1}\|_A \notag
\end{align}
where $\bm A\!=\!\text{diag}(a_n\!>\!0, n=1,\ldots, I\!\cdot\! J)$ is a positive definite matrix where each element $a_n$, $n\!=\!i\cdot j$ models the delay when UE $i$ is associated (de-associated) to (from) BS $j$, and $\|\cdot\|_A$ is its induced norm, i.e., $\|\bm x\|_A^2\!=\! \sum_{n}\! a_n x_n^2$ and  $\|\bm x\|_{A*}^2\!=\!\sum_{n}\!  x_n^2/a_n$ 
\cite{beck-book}. For instance, for $I\!=\!3$ and $J\!=\!2$, we have: $\|\bm x_t \!-\! \bm x_{t-1}\|_A^2=\sum_{i=1}^3\sum_{j=1}^2a_{ij}(x_{ij}(t)\!-\!x_{ij}(t\!-\!1))^2$; if UE $i\!=\!1$ moves from BS $j\!=\!1$ to $j\!=\!2$, it yields a total HO delay $a_{11}\!+\!a_{12}$, where $a_{11}$ ($a_{12}$) is the delay due to the de-association (association) from (to) $j\!=\!1$ ($j\!=\!2$). The scalarization parameter $\gamma$ is used to normalize units and prioritize one criterion (i.e., throughput) over the other (i.e., HO delays), based on the preferences of each operator.

This HO model departs from previous works, e.g., \cite{andrews-association, kelleler-jsac23} and references therein, that merely count the association changes as if all HOs had the same impact on performance. Clearly, the analysis in Sec. \ref{sec:data-analysis} showed this assumption not to be accurate in today's heterogeneous networks. Hence, we opt here instead to modulate the HO costs with parameters reflecting the potentially-different HO time for each UE-BS pair, and additionally with the tunable $\gamma$ weight. In particular, based on Sec. \ref{sec:data-analysis}, we will be using in our numerical evaluations the BS RAT and UE type as the main features for the HO delays. Finally, we define the decision set: 
\begin{align*}
	&\mathcal X=\bigg\{ \bm x \in \{0,\!1\}^{I\cdot J}  \  \Big |   \sum_{j\in \mathcal J}{x}_{ij}=1,  i\in \mathcal I \bigg\},
\end{align*}
and its convex hull $\mathcal X^{c}\!=\!\text{co}(\mathcal X)$ that relaxes the integrality, i.e., $\bm x \! \in \! [0,1]^{I\cdot J}$. We will use $\bm x$ when referring to the discrete associations, and denote with $\bm x^m \!\in\! \mathcal{X}^c$ the respective relaxed vector. Putting these together, we have the next result. 

\vspace{1mm} 
\noindent \textit{Proposition 1.} The throughput and HO cost function is concave on $\mathcal X^c$: $f_t(\bm x\big)\!\triangleq\!g_t(\bm x)+h(\bm x, \bm x_{t-1})=$
\begin{align*}
\sum_{i\in \mathcal I}\sum_{j\in \mathcal J}\! x_{ij}(t)\log c_{ij}(t) - \sum_{j\in \mathcal J} y_{j}(t)\log y_{j}(t)-\gamma\|\bm x-\bm x_{t-1}\|_A.
\end{align*} 
The concavity of $g_t(\bm x)$ for $\bm x \in \mathcal X^c$ is proved in \cite{andrews-association}, and it follows that subtracting the A-norm preserves the property.

The performance of the algorithm will be assessed using the \emph{Expected Dynamic Regret}, defined as:
\begin{align}
\!	\E\left[{\mathcal R}_T\right]\!\triangleq \!\sum_{t=1}^Tf_t(\bm x_t^{\star}) \! -  \sum_{t=1}^T	\E\Big[ f_t(\bm{x}_t)\Big], \label{regret-metric}
\end{align}
where $\{\bm{x}_t\}_t$ are the algorithm decisions, $\{\bm x_t^{\star}\}_t$ is the benchmark, and the expectation captures any randomization in the algorithm. Our goal is to design an algorithm that ensures this gap diminishes with time, $\lim_{T\rightarrow \infty} \E[{\mathcal R}_T]/T\!=\!0$ for any possible benchmark sequence $\{\bm x_t^{\star}\}_t$. In other words, we compare our algorithm with the best oracle that has full information at $t\!=\!1$ for the  SINR and HO delays, for all users and BSs, for the entire horizon $T$. Clearly, this is a very competitive benchmark, going beyond static and (per-slot) dynamic regret (see discussion in \cite{bianchi-switching}), and thus a sublinear-regret algorithm in this context is highly desirable.

\begin{figure}[t]
	\centering
	{\includegraphics[width=0.85\columnwidth]{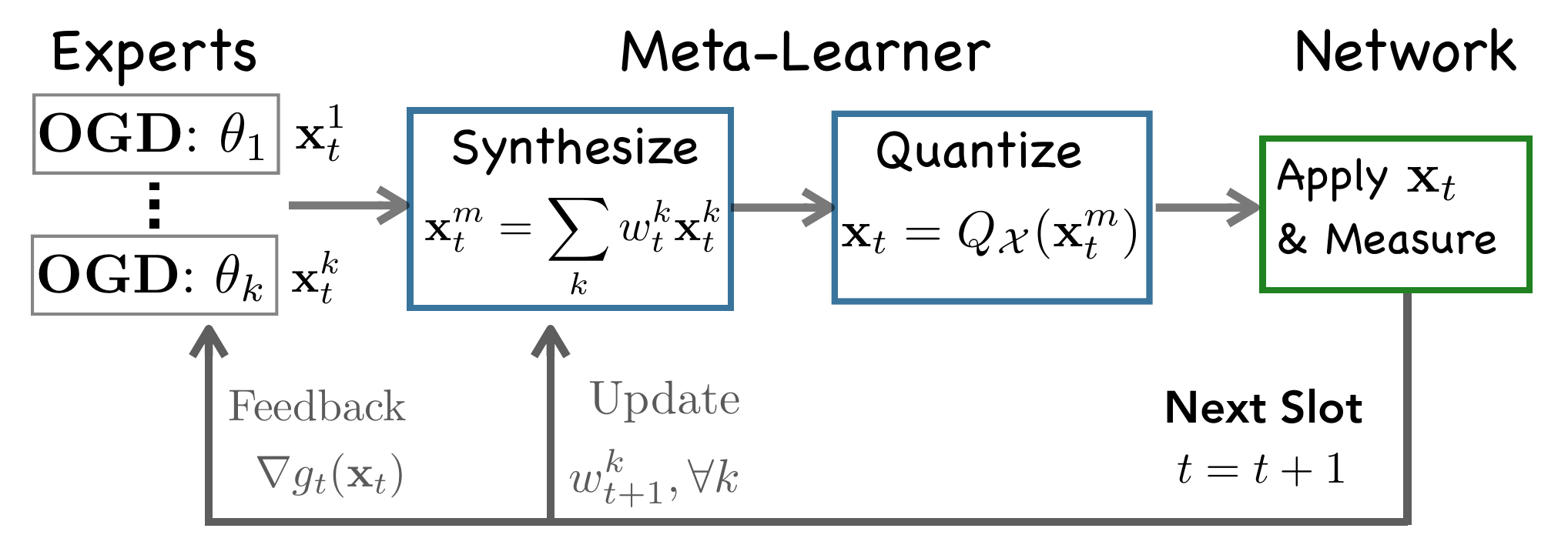}}
        \vspace{-2mm}
	\caption{\small{Learning mechanism.}}
	\label{Fig:learning-mechanism}
        \vspace{-5mm}
\end{figure}

\noindent\textbf{Online Algorithm}. The proposed learning mechanism is summarized in Fig. \ref{Fig:learning-mechanism}. There is a meta-learner that receives suggestions for the association policy from $K$ agents (the \emph{experts}); creates accordingly a weighted policy; and learns gradually how much to trust each expert based on its performance. In turn, each expert learns dynamically the best association strategy, based on feedback from the meta-learner. The experts use different learning steps $\bm \theta=(\theta_k, k\in\mathcal K)$, that are selected so as to ensure at least one of them will perform optimally w.r.t. the yet-to-be-encountered problem conditions. The details of the mechanism are provided in Algorithm 1 (\texttt{LDA}). At the beginning of each slot, each expert $k$ shares its suggestion $\bm{x}_t^k$, and the meta-learner synthesizes them:
\begin{align}
\bm x_t^m=\sum_{k\in\mathcal K}w_t^k\bm{x}_t^k, \label{eq:meta-mix}
\end{align}
where the weights $\bm{w}_t\!=\!(w_t^k, k\in \mathcal K)$, with $\bm w_t^\top \bm{1}_K\!=\!1$, is what the meta-learner needs to learn. It follows that if $\bm{x}_t^k\!\in\!\mathcal X^c, \forall k$, then $\bm{x}_t\!\in\!\mathcal X^c$. Next, the meta-learner creates a binary decision vector $\bm{x}_t \!\in\! \mathcal X$ (so as to be implementable) using the quantization routine $Q_{{\mathcal X}}$. For this operation, one can use any unbiased sampling technique, as long as $\E[\bm x_t]\!=\!\bm x_t^m$. {For instance, Madow's sampling \cite{madow} ensures this condition which, for the structure of $\mathcal X$, simply picks an element from $\bm{x}_i(t)$ with probabilities $\bm{x}_i^m(t)$, for each $i\!\in\!\mathcal I$.}

At the end of the slot, the controller observes the system parameters\footnote{Since UEs report their channel gains with all reachable base stations (and not only with the one they are associated), we have a full-information model.} and calculates the gradient $\nabla g_t(\bm x_t)$, which is sent to all experts.  Then, the meta-learner updates the weights:
\begin{align}\label{experts-weights-update}
	w_{t+1}^k=\frac{w_{t}^ke^{\beta\ell_t(\bm x_t^k)}}{\sum_{k\in\mathcal K} w_{t}^ke^{\beta \ell_t(\bm x_t^k)} }
\end{align}
using the surrogate (i.e., partially linearized) loss:
\begin{align}\label{loss-function}
\ell_t(\bm x_t^k)=\dtp{\nabla g_t(\bm x_t) }{\bm x_t^k - \bm x_t} -\gamma \|\bm{x}_t^k - \bm{x}_{t-1}^k\|_A.
\end{align}
It is interesting to observe that all experts use the same gradient and not the gradient of their own action. This is possible due to the concavity of $g_t$, which can be upper bounded by the linearized loss at \emph{any} information point; see also \cite{zhang-18}.

\begin{algorithm}[t] 
	{\small{	
			\nl \textbf{Input}: Horizon $T$; $K$ experts with steps $\{\theta_k\}_{k\in \mathcal K}$; Meta-learning step $\beta$\\ 
			\nl \textbf{Initialize}: set ${w}_1^{k}=\frac{1+(1/K)}{k(k+1)}, \forall k\in \mathcal K$; draw $\bm x_1^k\in \mathcal X, \forall k\in \mathcal K$; sort $\theta_1\leq \theta_2\leq \ldots \leq \theta_K$; \\
			\nl \For{ $t=1,2,\ldots, T$  }{
				\nl Each expert $k\in \mathcal K$ shares its decision $\bm x_t^k$;\\
				\nl  The controller synthesizes  $\bm x_t^m$ using \eqref{eq:meta-mix};\\
				\nl  The controller implements $\bm{x}_t=Q_{{\mathcal X}}(\bm x_t^m)$;\\		
				\nl The controller sends $\nabla g_t(\bm x_t)$ to experts; \\
				\nl The controller updates its weights using \eqref{experts-weights-update}; \\							
				\nl Each expert updates its decision using \eqref{expert-update}; \\							
			}
			\caption{{\small{Learning Dynamic Associations (\texttt{LDA})}}}\label{alg1}
	}}	
	\setlength{\intextsep}{0pt} 
\end{algorithm} 

Finally, the experts perform an online gradient ascent:
\begin{align}\label{expert-update}
	\bm{x}_{t+1}^k=\Pi_{\mathcal X}\Big( \bm{x}_t^k + \theta_k\nabla g_t(\bm x_t)\Big),
\end{align}
so as to produce their next suggestion. The number of experts, their rates, initial weights $\bm w_1$, and meta-learning step $\beta$, are selected to ensure the regret convergence, as explained next.

\noindent\textbf{Performance Analysis}. We first introduce some key parameters that are used below. Defining $a_{\max}=\max_{n\leq I\cdot J}{a_n}$ we obtain the bounds: 
\begin{itemize}
	\item $\|\bm x\!-\!\bm x'\|_2 \!\leq\! D$, $\|\bm x\!-\!\bm x'\|_A \!\leq\! \sqrt{a_{max}} D\!\triangleq \! D_A$, $\|\bm x\!-\!\bm x'\|_{A*} \!\leq\! D/\sqrt{a_{max}}\!\triangleq \! D_{A*}$ with $D=\sqrt{2I}$.
	\item $\|\nabla g_t(\bm x)\|_2 \!\leq\! G$, $\|\nabla g_t(\bm x)\|_A \!\leq \!\sqrt{a_{max}} G\!\triangleq\!G_A$, with $G=\sqrt{ IJ(\log J+1/\ln 10)^2}$.
\end{itemize}

To characterize the performance of Algorithm \texttt{LDA}, we proceed in two steps. First, we bound the regret of the relaxed (continuous) performance and switching cost, with the following Lemma that we prove in the appendix:

\noindent\textit{Lemma 1.}  Using the following parameters:
\begin{itemize}
	\item	$K=\left\lceil\log_2\sqrt{1+\!2T} \right \rceil+1$.

	\item	$\theta_k=2^{k-1}\Big[\frac{D_A^2}{T(G^2+2G_A)}\Big]^{1/2}$,  $k=1,\ldots,K$.
				
	\item	$\beta=1/\sqrt{T\nu}$, with $\nu\triangleq (2GD\!+\!D_A)^2(D_A\!+\!(1/8))$.				
\end{itemize}	
The continuous decisions $\{\bm x_t^m\}_t$, where $\bm x_t^m\in \mathcal{X}^c$ ensure:
\begin{align}
\!\!\sumT f_t(\bm{x}_t^{\star}) - \sumT f_t(\bm{x}_t^m) \!\leq\! \mathcal{O}\left( \sqrt{T(1+P_T)}	\right)
\end{align}
with the benchmark's total HO delay $P_T\!=\!\sumT \!\!\|\bm{x}_t^{\star}\!-\bm{x}_{t-1}^{\star}\|_A$.

As is expected in dynamic regret, the bound depends on the variability of the benchmark which here, interestingly, has the physical meaning of HO delays. In any case, the meta-learner follows the benchmark and gradually decreases the gap. These decisions refer to a continuous-valued association strategy ($\bm{x}_t^m\!\in\!\mathcal X^c$) similarly to \cite{andrews-association, kelleler-jsac23} and others,  and can be interpreted as probabilistic associations. Unlike these prior works, here we make an extra step to provably bound the expected dynamic regret of the implementable \emph{discrete} associations. The observation we utilize is that $\bm 0\preceq \bm x_t,\bm x_t^m\preceq \bm 1$, and are related through an unbiased sampling, i.e., $\E[\bm x_t]\!=\!\bm x_t^m$. Based on this, we are able to obtain the following result, proved in the appendix.

\noindent\textit{Theorem 1}. Algorithm \texttt{LDA} ensures the following bound against \emph{any} benchmark sequence $\{\bm{x}_t^*\}_{t=1}^T$:
\begin{align}
\E\left[{\mathcal R}_T \right]\! \leq \! \mathcal{O}\left(\! \sqrt{T(1\!+\! P_T)} \right) \!+\! G_fT\sqrt{I\!-\!(I/J) } \notag
\end{align}
where $\|f_t(\bm x)\|_2\!\leq\!G_f, \forall \bm x\in \mathcal X, \forall t$.

\noindent\textbf{Discussion}. The theorem shows that the regret of \texttt{LDA} is sublinearly dependent on the oracle's HO delay, which marks its learning capability. There is also an unavoidable non-diminishing error term due to discretization. In fact, this error can be eliminated  by normalizing properly the step $\beta$ and using the doubling trick. Due to lack of space we kindly refer the reader to \cite[Lemma 3]{lesage-tac-21} for further details. The doubling trick will also eliminate the need to know in advance the horizon $T$. Besides, even without the step normalization, we find the error to be only 1.1--1.3\% of 
the objective value, when tested on static and volatile scenarios (see Sec. \ref{sec:evaluation}).

Moreover, it is important to stress that Algorithm \texttt{LDA} is oblivious to information such as the SINR during the slot, which, in practice, is unknown at the beginning of each slot \cite{kalntis_tcom24}. This is a key difference of the proposed approach compared to prior works such as \cite{andrews-association, kelleler-jsac23}. Further, \texttt{LDA} is scalable to the number of UEs and BSs and is amenable to near-real-time execution as it has relatively lightweight operations and thus, can be implemented in O-RAN \cite{kalntis22, kalntis_tcom24}.

\section{Extensions}\label{sec:exensions}

Finally, we discuss two key extensions: \emph{(i)} when HO delays are unknown, where we show that \texttt{LDA} can adapt to them dynamically; and \emph{(ii)} when there is an ML mechanism that proposes associations based on SINR/mobility forecasts \cite{andrews-association, tassiulas-globecom}, and prove that \texttt{LDA} can seamlessly benefit from them.

\noindent\textbf{Time-varying HO delays}. Algorithm \texttt{LDA} does not require knowing in advance the elements of $A$, i.e., the HO delays. To see this, first, observe that the experts do not use the HO delay when they update their decisions (only the throughput $\nabla g_t(\bm x_t)$). And secondly, the meta-learner uses the HO delays when it calculates $\ell_t(\bm x_t^k), \forall k\in \mathcal K$ to update the weights $\bm w_t$, which takes place \emph{after} the HOs are realized. This flexibility allows to tackle cases where the HO delays for each type of UEs and BSs, change with time. From a technical point of view, instead of the fixed $A$-norm, \texttt{LDA} can use a time-varying norm $\|\cdot\|_{A_t}$, where each $A_t$ captures the delay that was observed (a posteriori) at each slot $t$. This information is then used to calculate the weights $\bm w_{t+1}$ and association $\bm x_{t+1}$. The proofs of Lemma 1 and Theorem 1 follow nearly verbatim, with the modification of changing the fixed norm to the time-varying norms, and redefining $P_T\!=\!\sumT\!\|\bm x_t^{\star} \!-\! \bm x_{t-1}^{\star}\|_{A_t}$.

\noindent\textbf{Encompassing Forecasters}. Unlike our approach that learns from runtime data, several recent works \cite{andrews-association, tassiulas-globecom} proposed to decide the HOs leveraging forecasters for SINRs. The output of such tools, let us denote it $\{\bm x_t^p\}_t$, can be very close to $\{\bm x_t^{\star}\}_t$ if the forecasters are accurate, but very suboptimal otherwise (e.g., if there is a distribution shift). Our framework, on the other hand, can benefit from such tools in a robust fashion, by assessing their accuracy in real time. The meta-learner can include $\{\bm x_t^p\}_t$ as the $(K\!+\!1)$th expert, and assess its performance in real-time so as to discard it if proved inaccurate. And if the forecasting tool is effective, the regret will improve significantly. This can be seen by revisiting the proof of Lemma 1 that utilizes the Hedge algorithm (see \cite[Lemma 1]{zhang-18}), which bounds the regret of the meta-learner from the best expert (thus, also from the forecaster $p$) as:
\begin{align*}
&\max_{k\in\mathcal K\cup {p}}\sumT \ell_t(\bm x_t^k)- \sumT \ell_t(\bm x_t^m) \stackrel{(\alpha)}= \\
&\sumT \ell_t(\bm x_t^p)- \sumT \ell_t(\bm x_t^m)\leq \frac{\beta T c^2}{8}+\frac{1}{\beta}\ln\frac{1}{w_{1}^k}
\end{align*}
where $c\!=\!2GD\!+\!D_A$ and $(\alpha)$ holds when the forecaster is the best expert. Defining its error $\sum_t \ell_t(\bm x_t^{\star})\!-\!\ell_t(\bm x_t^p)\!=\! \epsilon_T$ we get:
\[
\E[\mathcal R_T]\leq \frac{\beta T c^2}{8}+\frac{1}{\beta}\ln\frac{1}{w_{1}^k} + \epsilon_T +G_fT\sqrt{I-(I/J)}
\]
which is optimized when $\beta\!=\!\mathcal O(1/\sqrt T)$ (as in Theorem 1). Comparing this result with Theorem 1, we see that when the forecaster is successful, the overall performance improves by dropping an entire term, and does not depend on $P_T$; while, when the forecaster is found to be inaccurate, \texttt{LDA} maintains the previous performance as it relies on a different expert. The idea of combining forecasters with online learning is often referred to as \emph{optimistic learning}~\cite{naram-tmc,rakhlin-optimistic}, and we apply it, for the first time, in the context of SOL with dynamic regret.

\section{Performance Evaluation}\label{sec:evaluation}

\begin{figure}[t]
    \centering
    \begin{subfigure}[H]{0.235\textwidth}
         \centering
         \includegraphics[width=\columnwidth]{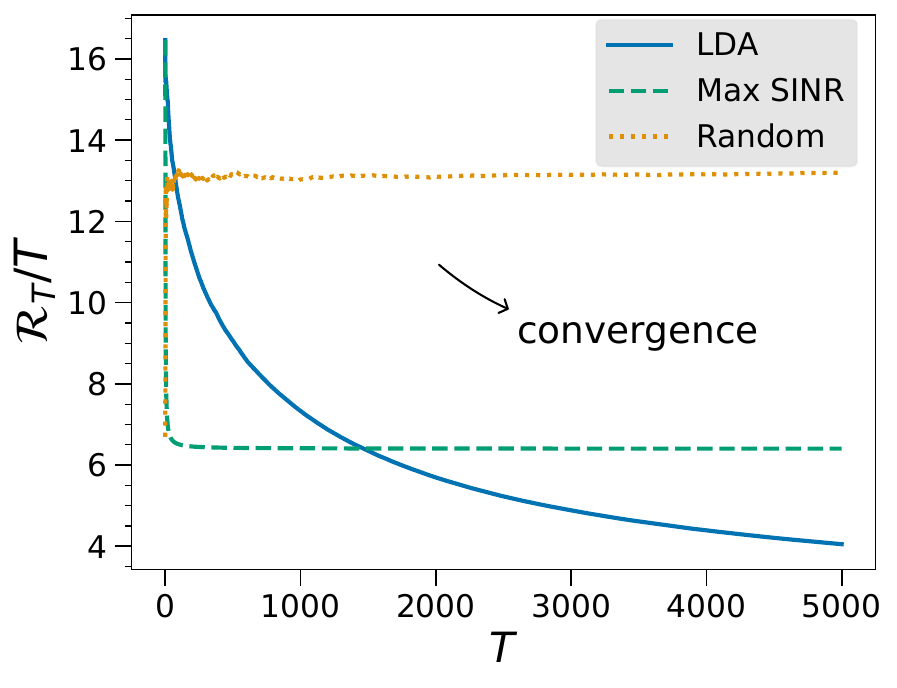}
         \caption{}
         \label{fig:static_regret}
     \end{subfigure}
     \begin{subfigure}[H]{0.235\textwidth}
         \centering
         \includegraphics[width=\columnwidth]{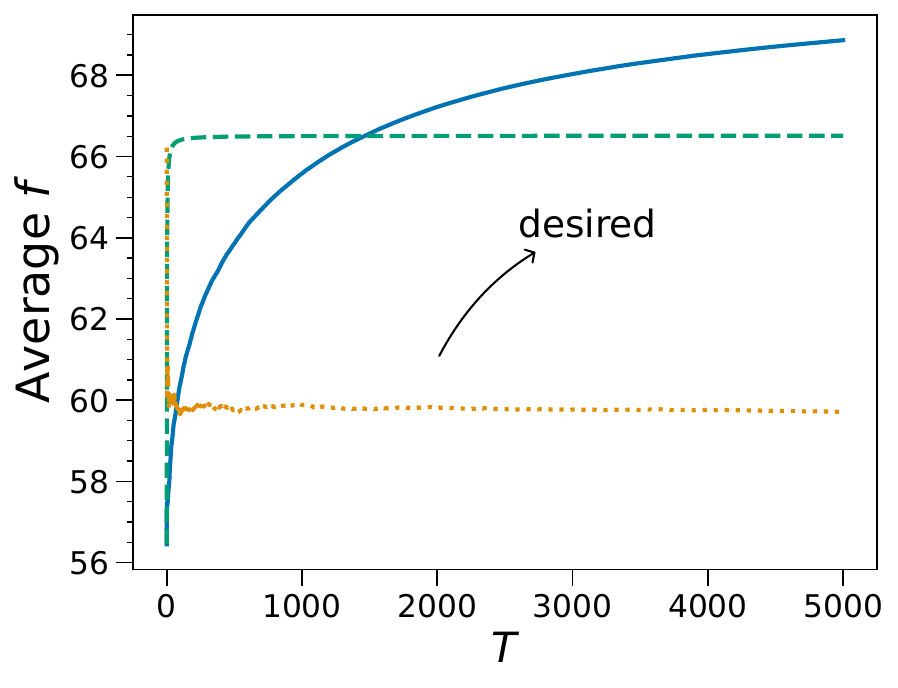}
         \caption{}
         \label{fig:static_average_f}
     \end{subfigure}
     \vspace{-2mm}
     \caption{\small{Static scenario for $\gamma=20$: (a) average dynamic regret and (b) average obtained objective function $f$.}}
     \label{fig:static_regret_average_f}
     \vspace{-6mm}
\end{figure}

\texttt{LDA} undergoes rigorous evaluation across multiple scenarios that encompass both real-world conditions (actual users, cells, and SINRs) and synthetic ones, verifying that its efficacy is broadly applicable. We utilize these scenarios to investigate the algorithm's learning convergence, and compare its performance with different benchmarks in terms of \emph{(i)} attained dynamic regret, \emph{(ii)} accumulated objective function, throughput, and HO cost, and \emph{(iii)} impact of incorporating $\bm A$ versus using a simple $L2$ norm, or none at all. The employed benchmarks are \emph{(i)} an \texttt{LDA}-based algorithm using the Euclidean norm in the HO cost, called \texttt{LDA 2-norm}, aiming to highlight the significance of the A-norm; \emph{(ii)} an advanced greedy algorithm, \texttt{Max SINR}, that assigns UEs to BSs based on the maximum SINR from the previous slot (the current slot's SINR is only known post-association), disregarding HO delays; \emph{(iii)} an optimal \texttt{Oracle}, used in the definition of dynamic regret in \eqref{regret-metric}, which has complete knowledge of future; \emph{(iv)} a basic \texttt{Random} algorithm that makes the associations randomly, serving as a minimal benchmark.

\begin{figure}[t]
    \centering
    \begin{subfigure}[H]{0.235\textwidth}
         \centering
         \includegraphics[width=\columnwidth]{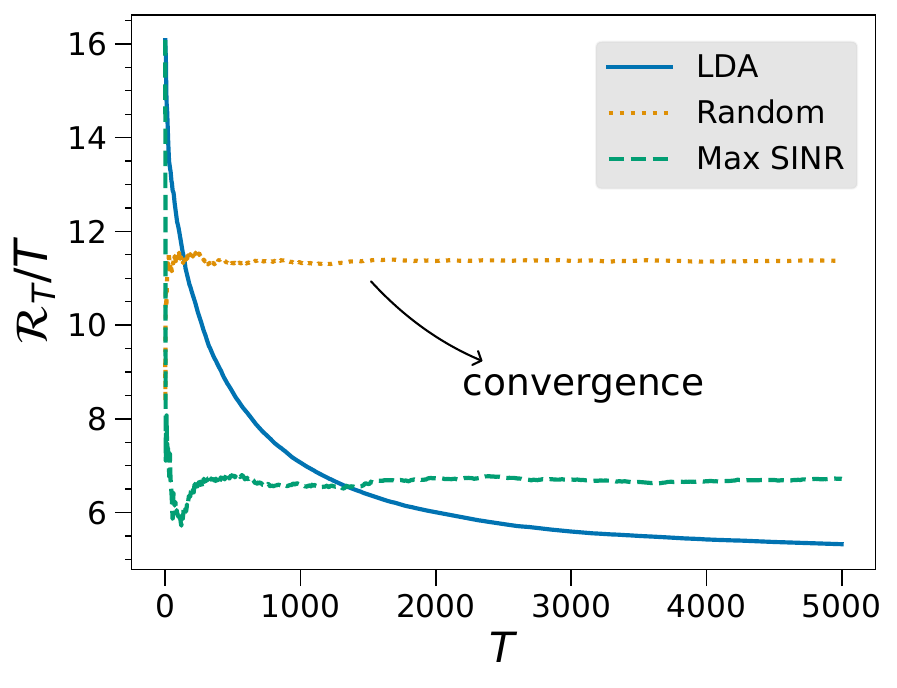}
         \caption{}
         \label{fig:volatile_regret}
     \end{subfigure}
     \begin{subfigure}[H]{0.235\textwidth}
         \centering
         \includegraphics[width=\columnwidth]{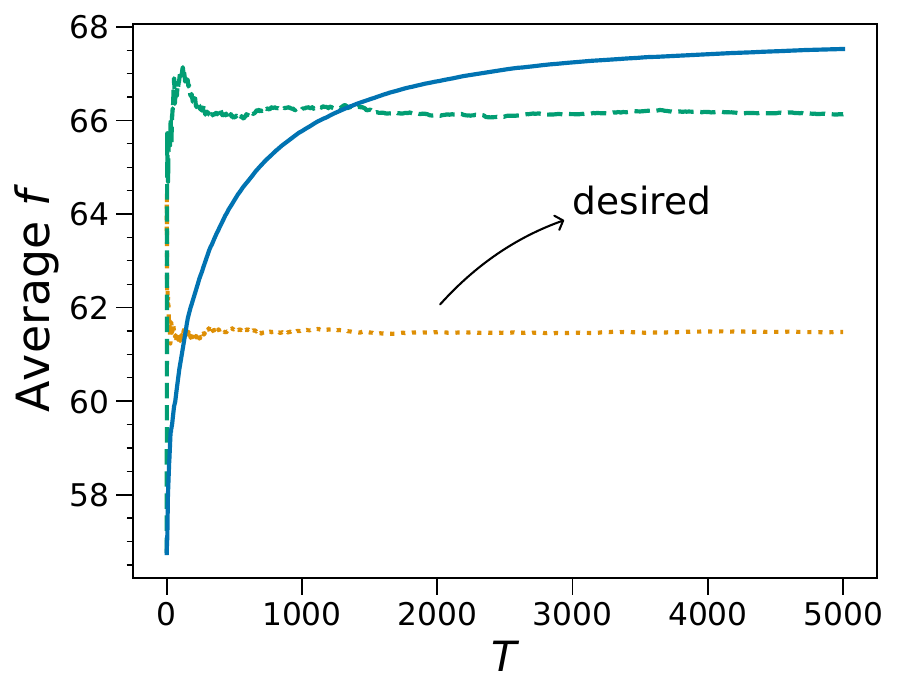}
         \caption{}
         \label{fig:volatile_average_f}
     \end{subfigure}
      \vspace{-2mm}
     \caption{\small{Volatile scenario for $\gamma=20$: (a) average dynamic regret and (b) average obtained objective function $f$.}}
     \label{fig:volatile_regret_average_f}
     \vspace{-5mm}
\end{figure}

\noindent\textbf{Synthetic scenarios}. We select two synthetic scenarios that are in line with those in prior work~\cite{ayala2021bayesian, kalntis22, kalntis_tcom24}, and which we define as follows, for $T=5k$ slots: ($i$) \textit{static}: the SINR $s_{ij}(t)$ remains constant across all slots, and ($ii$) \textit{volatile}: $s_{ij}(t)$ fluctuates every 5 slots within the range of [10, 30]dB \cite{3gpp_36_133}, encompassing poor to excellent values. In both scenarios, we randomly select bandwidths $w_j \in \{5, 10, 15, 20\}$ MHz \cite{3gpp_36_101}, while $\bm A$ takes random values within [0, 1], and $\gamma=20$ (penalizes more HO cost, without sacrificing throughput). We choose a smaller case study with $I\!=\!100$ UEs and $J\!=\!10$ BS to facilitate the calculation of average regret, as determining the best oracle is computationally intensive. However, it is important to note that the best oracle is not necessary for running \texttt{LDA}. Figs. \ref{fig:static_regret_average_f} and \ref{fig:volatile_regret_average_f} show that \texttt{LDA} identifies the best benchmark (diminishing regret) just in solely $T\!=\!5k$ slots. This verifies our claim (see Sec. \ref{sec:algorithm}) that the error due to discretization is small. Conversely, \texttt{Max SINR} fails to converge, even though the SINR does not change, with the regret remaining constant at 6.5 (exploiting sub-optimal associations). This trend is also evident in the average objective function, where \texttt{LDA} outperforms \texttt{Max SINR} after $t=1.4k$. As anticipated, \texttt{Random} algorithm performs poorly.

\noindent\textbf{Real-world scenarios}. To test \texttt{LDA} in real-world scenarios, we use crowdsourced measurements that contain various signal metrics (e.g., SINR and SNR) along with their precise latitude and longitude information and the concerned BS, at every second. This dataset presents a highly competitive scenario as SINRs can vary arbitrarily, and occasionally exhibit adversarial behavior. The crowdsourced measurements are anonymized, hence we lack specific UE information (e.g., its type or exact mobility pattern). Therefore, $\bm A$ is chosen from the distribution of our data, as shown in Figs. \ref{fig:ue_diversity}, \ref{fig:RAT_ho_delays}, and \ref{fig:ue_ho_delays}, ensuring a variety of UE types and HO delays are considered. Moreover, we adopt the Gauss-Markov mobility model \cite{gauss_markov}, in line with prior works~\cite{andrews-association}, with randomness parameter $a\!=\!0.5$ ($a\!=\!0$ being totally random and $a\!=\!1$ being linear motion) and consider velocities in [1, 28] meters per second, and velocity variances in [0, 14]; thus having from pedestrians to fast-moving UEs. 

\begin{figure}[t]
    \centering
    \begin{subfigure}[H]{0.235\textwidth}
         \centering
         \includegraphics[width=\columnwidth]{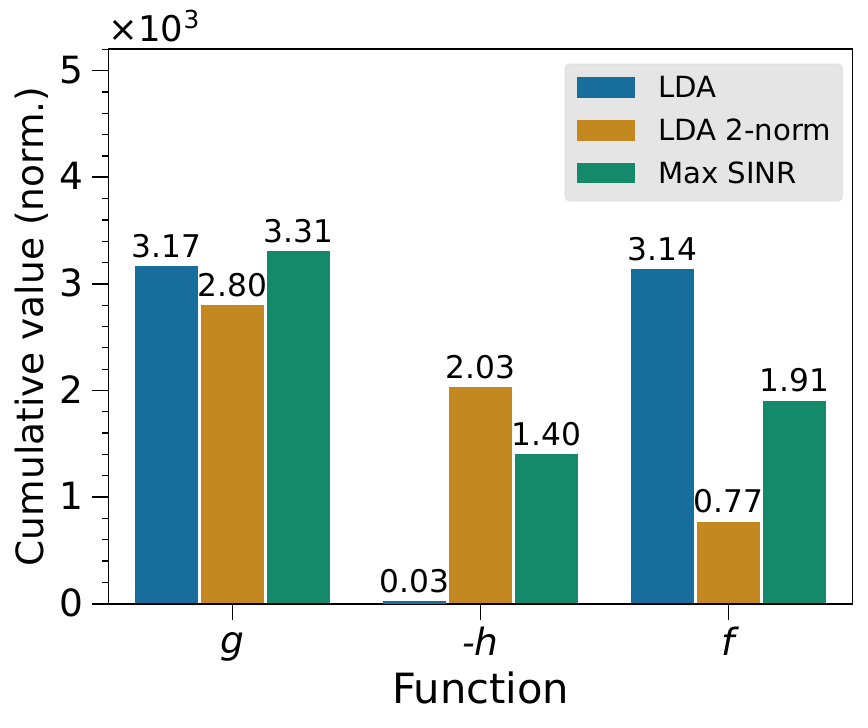}
         \caption{}
         \label{fig:real-A-small}
     \end{subfigure}
     \begin{subfigure}[H]{0.235\textwidth}
         \centering
         \includegraphics[scale=0.29]{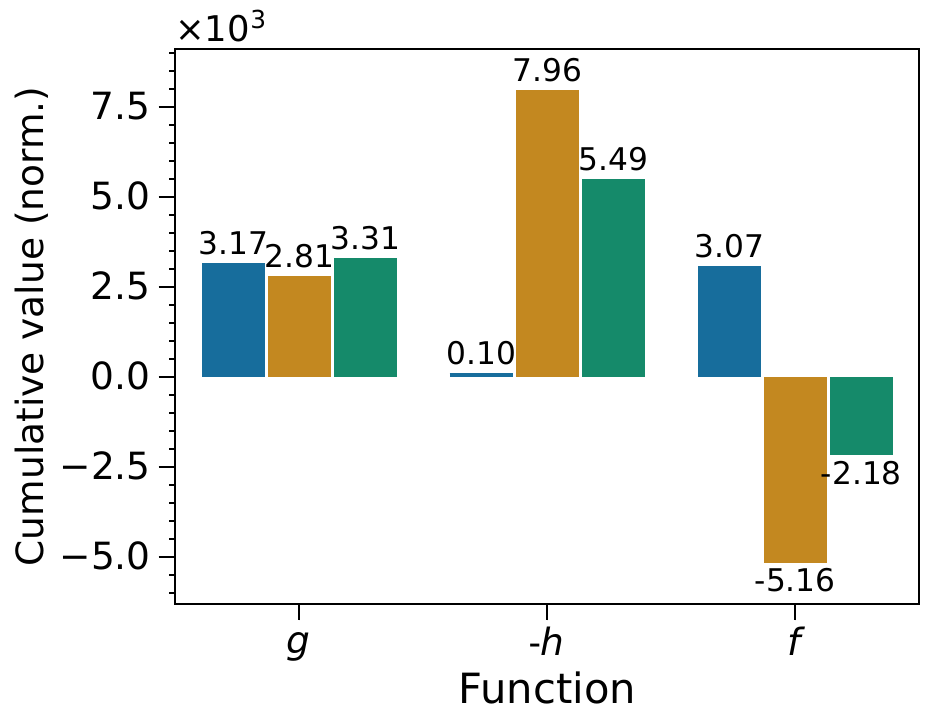}
         \caption{}
         \label{fig:real-A-big}
     \end{subfigure}
     \vspace{-1mm}
     \caption{\small{Comparison of throughput ($g$), negative HO cost ($-h$), and total ($f$) values for the real case, when (a) $\gamma=5$, (b) $\gamma=20$.}}
     \label{fig:real-case}
     \vspace{-6mm}
\end{figure}

At $t=1$, 1k UEs are placed randomly on the map at a different location with recorded SINR measurements. At each subsequent slot $t=2, ..., T$, for $T=10k$, each user moves in accordance with their individual parameters, to their own new location. To incorporate the signal measurements, we map each new location to the nearest location with available measurements, and continue for the next slot similarly. We utilize measurements from an urban district and choose 25 of the BS with the most data. For the bandwidths of each BS, we use the real ones from our dataset.

We consider two different cases for $\gamma$: ($i$) $\gamma=5$, and ($ii$) $\gamma=20$ (chosen in accordance with throughput and HO cost values). Both cases prioritize high throughput, but the latter penalizes HO costs more. From Fig. \ref{fig:real-case}, we observe the cumulative $g$ values (normalized by the maximum value) are the same for each algorithm, in both $\gamma=5$ and $\gamma=20$ case; ensuring no throughput is sacrificed in turn of lower HO delays. At the same time, \texttt{LDA 2-norm} which does not consider the HO delays but solely the number of HOs, achieves $\times$67.7 and $\times$79.6 higher HO cost than \texttt{LDA} for $\gamma=5$ and $\gamma=20$, respectively. Similarly, although \texttt{Max SINR} achieves 4.4\% higher throughput, it does not take into account the HO costs, resulting in 39\% (171\%) lower total values for $\gamma=5$ ($\gamma=20$). Therefore, \texttt{LDA} performs optimally in terms of HO costs, without sacrificing throughput.

\begin{figure}[t]
	\centering
	{\includegraphics[width=0.85\columnwidth]{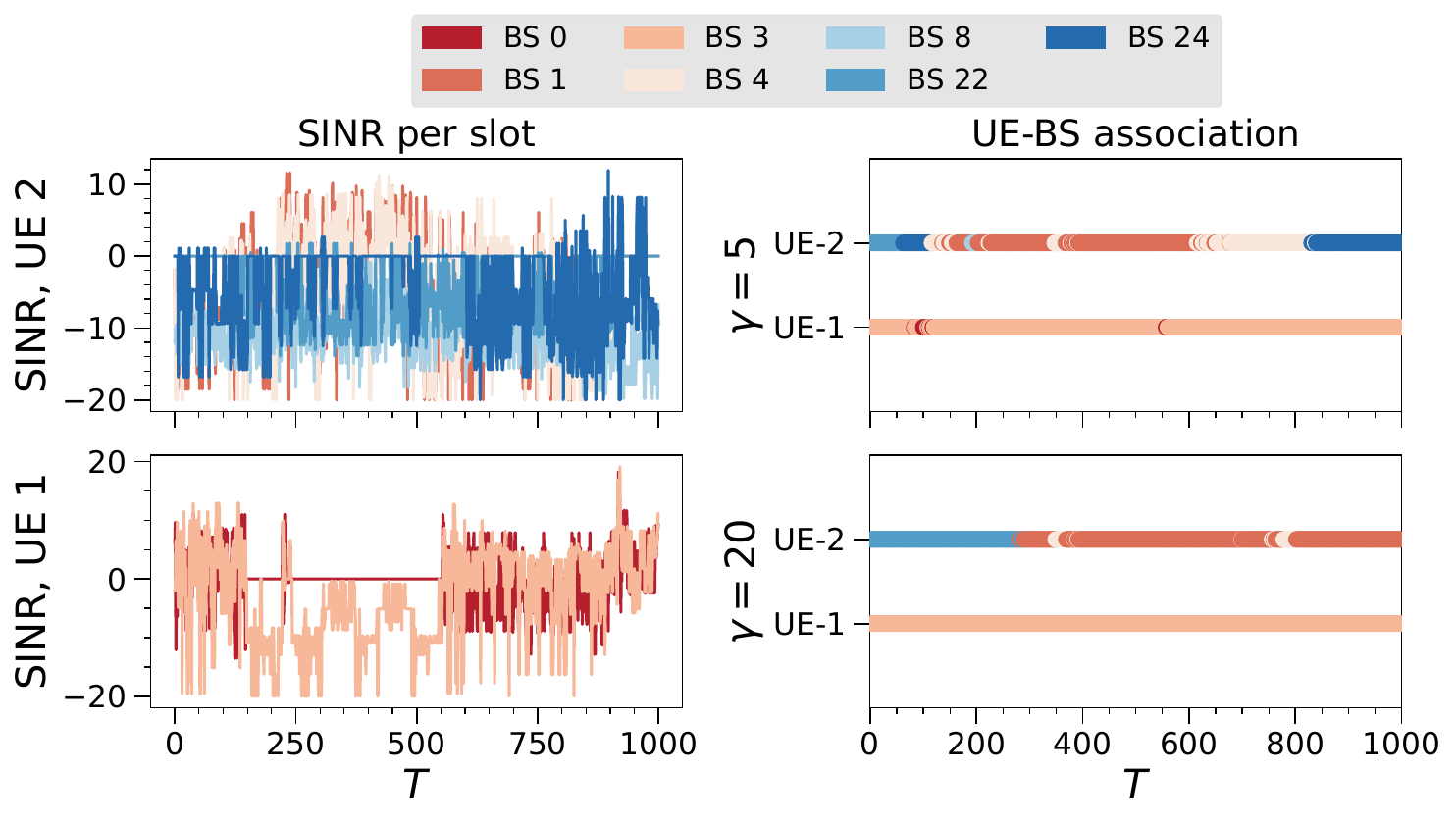}}
    \vspace{-2mm}
	\caption{\small{SINR per slot of two real users (upper and lower left) for $T=1k$, and the UE-BS associations that \texttt{LDA} chooses, for $\gamma=5$ (upper right) and $\gamma=20$ (lower right).}}
	\label{fig:real-sinr}
 \vspace{-7mm}
\end{figure}

To further deep-dive into how \texttt{LDA} responds to $\gamma$, we show in Fig. \ref{fig:real-sinr} the SINR of two UEs for the first $t=1k$ slots, and the UE-BS associations for $\gamma=5$ and $\gamma=20$. In the former case, we observe that both users change BS more frequently, e.g., for $\gamma=5$, UE-1 changes 5 times between BS-3 and BS-0 even though the SINR of the latter becomes for a few slots slightly better than the former, while for $\gamma=20$, UE-1 stays constantly to BS-3. These results show that \texttt{LDA} can be easily adjusted by changing $\gamma$, to react more to SINR changes (i.e., maximize throughput), or take more into account the HO cost. The latter is a crucial aspect in mobile networks, where HOs may lead to the ping-pong effect \cite{pp_ho_2023}.

\vspace{-0.5mm}
\section{Conclusions}\label{sec:conclusions}

This study addresses the problem of handover (HO) optimization in cellular networks using Smooth Online Learning (SOL). Analyzing countrywide data from a European mobile network with over 40M users and 370k sectors, we identify key correlations between HO failures/delays and the characteristics of radio cells and devices. We develop a dynamic model for UE-to-cell associations that incorporates device and cell features with minimal assumptions. Our proposed algorithm, \texttt{LDA}, aligned with the O-RAN paradigm, demonstrates robust performance and significant improvements in real-world and synthetic scenarios. These insights and solutions, based on the MNO’s perspective, are crucial for understanding and developing new mechanisms for smooth HOs, providing a foundation for future advancements in network performance.

\vspace{-2mm}
\section*{Acknowledgments}
This work has been supported by ($i$) the Spanish Ministry of Economic Affairs and Digital Transformation and the European Union – NextGeneration EU, in the framework of the Recovery Plan, Transformation and Resilience (PRTR) through the UNICO I+D 5G SORUS project, and UNICO I+D 5G 2021 refs. number TSI-063000-2021-3, TSI-063000-2021-38, and TSI-063000-2021-52, ($ii$)  the National Growth Fund through the Dutch 6G flagship project ``Future Network Services'', and ($iii$) the European Commission through Grant No. 101139270 (ORIGAMI) and 101192462 (FLECON-6G).

\vspace{-6mm}
\appendix

\noindent\textbf{Proof of Lemma 1}. The proof follows the rationale in \cite[Th. 4]{zhang-smoothed-ol}, \cite[Th. 3]{zhang-18}. First, we bound the regret of each expert w.r.t. benchmark, and then the regret of the meta-learner w.r.t. any expert. W.l.o.g., we set $\gamma=1$ and define:
\begin{align*}
\!s_t(\bm x)\!=\!\dtp{\nabla g_t(\bm x_t)}{\bm x\!-\!\bm x_t}, \text{ and} \ \bar{\bm x}_{t+1}^k\!=\!\bm x_{t}^k \!+\!\theta_k\nabla g_t(\bm x_t). 
\end{align*}
Then, we can write: $s_t(\bm{x}_t^{\star})-s_t(\bm x_t^k)=\dtp{\nabla g_t(\bm x_t)}{ \bm{x}_t^{\star}-\bm x_t^k }$
\begin{align*}
&\stackrel{(\alpha)}{\leq}\!\frac{1}{2\theta_k}\left( \|\bm{x}_t^{\star} \! - \!  \bm x_t^k \|_2^2 \! - \! \|\bar{\bm x}_{t+1}^k - \bm{x}_t^{\star}\|_2^2	\right) + \frac{\theta_k}{2}G^2\\ 
&+\|{\bm x}_{t+1}^k - \bm{x}_{t+1}^{\star}\|_2^2- \|{\bm x}_{t+1}^k - \bm{x}_t^{\star}\|_2^2) + \frac{\theta_k}{2}G^2\\
&\stackrel{(\beta)}=\!\frac{1}{2\theta_k}\big( \|\bm x_t^k - \bm{x}_t^{\star}\|_2^2\! -\! \|{\bm x}_{t+1}^k - \bm{x}_{t+1}^{\star}\|_2^2+ \\
&+(\bm{x}_{t+1}^k \! -\! \bm{x}_{t+1}^{\star}+ \bm{x}_{t+1}^k \!-\! \bm{x}_t^{\star})^\top(\bm{x}_t^{\star} - \bm{x}_{t+1}^{\star})\big) + \frac{\theta_k}{2}G^2\\
&\stackrel{(\gamma)}\leq\frac{1}{2\theta_k}\left( \|\bm x_t^k - \bm{x}_t^{\star}\|_2^2 \!-\! \|\bm{x}_{t+1}^k \! - \! \bm{x}_{t+1}^{\star}\|_2^2\right)+ \\
&+\frac{D_{A*}}{\theta_k}\|\bm{x}_t^{\star} \!-\! \bm{x}_{t+1}^{\star}\|_A \!+\! \frac{\theta_k}{2}G^2.
\end{align*}
$(\alpha)$ uses the identity $\dtp{x}{y}\!=\!(\|x\|_2^2 \!+\! \|y\|_2^2 \!-\! \|x\!-\!y\|_2^2)/2$, the definition of $\bar{\bm x}_{t+1}^k$ and  $\|\nabla g_t(\bm x_t)\|_2\!\leq \! G$; $(\beta)$ uses the projection non-expansiveness, adds/subtracts $\|\bm{x}_{t+1}^k \!-\! \bm{x}_{t+1}^{\star}\|$ and $\|x\|_2^2 \!-\! \|y\|_2^2 \!=\! (x\!-\!y)^\top(x\!+\!y)$; and $(\gamma)$ uses Cauchy-Schwartz, triangle inequality, and $D_A*$. Telescoping to $T$ and using $\|\bm x_1^k \!-\! \bm{x}_1^{\star}\|_A^2\!\leq\! D_A$, gives
\begin{align*}
\!\!\sum_{t=1}^T\!s_t(\bm{x}_t^{\star})\!-\!s_t(\bm x_t^k)\!\leq\! \frac{D_A}{2\theta_k} \!+\! \frac{D_{A*}P_T}{\theta_k}\!+\! \frac{\theta_kTG^2}{2}.
\end{align*}
Next, we bound the switching cost of (any) expert $k$:
\vspace{-2mm}
\begin{align*}
	&\sumT \|\bm x_t^k - \bm x_{t-1}^k\|_A=\sum_{t=0}^{T-1}\|\bm x_{t+1}^k-\bm x_t^k\|_A \leq \\ 
	&\sum_{t=0}^{T-1}\| \bar{\bm x}_{t+1}^k - \bm x_t^k \|_A=\sum_{t=0}^{T-1}\|-\theta_k\nabla g_t(\bm x_t)\|_A\leq \theta_kT G_A,
\end{align*}
and combining with the previous result, we get:
\vspace{-1mm}
\begin{align} 
\!&\sumT s_t(\bm{x}_t^{\star}) -\sumT\Big(	s_t(\bm x_t^k) -\|\bm x_t^k \!-\! \bm x_{t-1}^k\|_A	\Big)\!\leq\! \\ &\frac{D_A^2}{2\theta_k}+\frac{D_{A*}P_T}{\theta_k} + \theta_kT\left(\frac{G^2}{2}+G_A\right).\label{eq:lemma3}
\end{align}	

Next, we bound the gap of the meta-learner from all experts. Following \cite[Lem. 3]{zhang-smoothed-ol} and using the $A$-norm:
\begin{align*}
&\|\bm x_t^m - \bm x_{t-1}^m\|_A = \Big\|  \sum_{k\in\mathcal K}w_t^k\bm x_t^k - \sum_{k\in\mathcal K}w_{t-1}^k\bm x_{t-1}^k \Big\|_A\\
&\leq  \Big\|  \sum_{k\in\mathcal K}w_t^k(\bm x_t^k-\bm x)- \sum_{k\in\mathcal K}w_{t}^k(\bm x_{t-1}^k-\bm x)\Big\|_A\\ &+\Big\|\sum_{k\in\mathcal K}w_{t}^k(\bm x_{t-1}^k-\bm x)- \sum_{k\in\mathcal K}w_{t-1}^k(\bm x_{t-1}^k-\bm x) \Big\|_A\\
&\leq \sum_{k} w_t^k \left\|\bm x_t^k - \bm x_{t-1}^k\right\|_A + |w_t^k - w_{t-1}^k| \left\|\bm x_{t-1}^k - \bm x\right\|_A\\
&=\sum_{k\in \mathcal K} w_t^k \left\|\bm x_t^k - \bm x_{t-1}^k\right\|_A+D_A\|\bm w_t - \bm w_{t-1}\|_1.
\end{align*}
The relative loss of meta-learner is: $\sumT \Big(s_t(\bm x_t^k) \!-\! \|\bm x_t^k-\bm x_{t-1}^k\|_A\Big)-\sumT \Big( s_t(\bm x_t^m) \!-\! \|\bm x_t^m\!-\bm x_{t-1}^m\|_A\Big)$
\begin{align*}
&\leq \sumT \Big(	\sum_k w_t^k\left\|\bm x_t^k - \bm x_{t-1}^k\right\|_A  +\dtp{\nabla g_t(\bm x_t)}{\bm x_t^k\!-\!\bm x_t} \!\\
&\!- \left\| \bm x_t^k \!-\! \bm x_{t-1}^k \right\|_A\Big) \!+\! D_A\|\bm w_t\!-\!\bm w_{t-1}\|_1 =\\
&\sumT\! \left( \sum_k w_t^k\ell_t(\bm x_t^k) - \ell_t(\bm x_t^k)\right) \!+\! D_A\sumT \|\bm w_t \!-\! \bm w_{t-1}\|_1.
\end{align*}
The first term is bounded noting that $d\!\leq\! \ell_t(\bm x)\!\leq \!d \!+\!c$, with $d\!=\!-DG-D_A$, $c\!=\!2GD\!+\!D_A$ and using the Hedge bound \cite[Th. 2.2]{bianchi-book} (see also \cite[Lem. 1]{zhang-18}):
\vspace{-1mm}
\[
\sumT \ell_t(\bm x_t^m)-\min_{k\in\mathcal K} \left( \sumT \ell_t(\bm x_t^k)+\frac{1}{\beta}\ln\frac{1}{w_{1}^k}	\right)\leq \frac{\beta T c^2}{8}.
\]
Using the value of $c$ and the definition of $\bm x_t^m$, we get:
\begin{align}\label{distance-from-expert}
&\sumT\left( \sum_k w_t^k \ell_t(\bm x_t^k) - \ell_t(\bm x_t^k)	\right) \leq \notag \\ 
&\frac{1}{\beta}\ln \frac{1}{w_1^k} + \frac{\beta T(2GD+D_A)^2}{8}.
\end{align}
Next, we use the strong convexity of entropic FTRL \cite[Lem. 7]{mcmahan-survey}, as the basis for Hedge,  to arrive at:
\begin{align*}
	\|\bm w_t\!-\!\bm w_{t-1}\|_1 \!\leq\! \beta \left\|\left[\ell_{t-1}(\bm x_t^k)\right]_{k}\right\|_{\infty} \!\leq \! \beta (GD\!+\!D_A).
\end{align*}
Combining the above, we prove that $\forall k\in \mathcal K$, it is:
\vspace{-1mm}
\begin{align} 
&\!\sumT\!s_t(\bm x_t^k) \!-\! \|\bm x_t^k \!-\! \bm x_{t-1}^k\|_A\!- \!s_t(\bm x_t^m) \!+\! \|\bm x_t^m \!-\! \bm x_{t-1}^m\|_A\!\leq \notag \\
&\frac{1}{\beta}\ln\frac{1}{w_1^k}\!+\! \beta T\underbrace{\left[ (2GD+D_A)^2(D_A+(1/8))\right]}_{\nu}. \label{eq:regret-aux1}
\end{align}
and set $\beta= 1/\sqrt{T\nu}$ to balance the RHS (omitting $\ln$). 

Finally, considering the max and min values of $P_T$, the expert step that minimizes the RHS of \eqref{eq:lemma3} lies in:
\begin{align}
	\sqrt{ \frac{D_A^2}{T(G^2+2G_A)} }\leq \theta^{\star}\leq \sqrt{ \frac{D_A^2+2D_{A*}D_AT}{T(G^2+2G_A)} }
\end{align}
and inspecting the experts' steps $\{\theta_k\}_{k}$ in Lemma 1, we see that at least one expert step lies in that range. Thus, we obtain:
\begin{align}\label{eq:detailed-bound}
	&\sumT f_t(\bm x_t^{\star})-f_t(\bm x_t^m)\leq \sqrt{T}\Big[ \sqrt{\nu}\left(1+\ln(1/w_1^k)\right)+\notag \\
	&(G^2+2G_A)^{1/2}(D_A^2+2D_{A*}P_T)^{1/2}	\Big].
\end{align}

\noindent\textbf{Proof of Theorem 1}. It holds:

\begin{align*}
	&\E[\mathcal R_T]\!\leq\! \E\bigg[ \sumT \Big( f_t(\bm{x}_t^{\star})\!-\!f_t(\bm x_t)\!+\!f_t(\bm{x}_t^m) \!-\!f_t(\bm{x}_t^m) \Big)\bigg] \\
	&\!=\! \E\bigg[ \sumT f_t(\bm{x}_t^m) -f_t(\bm x_t) \bigg] \!+\!\sumT \Big(f_t(\bm{x}_t^{\star}) - f_t(\bm{x}_t^m)\Big).			
\end{align*}
The second term is bounded by Lemma 1. The first term is the discretization  \emph{error}, and we bound it as follows: 
\begin{align*}
	&\!\E\left[ \!\sumT\! \Big(f_t(\bm{x}_t^m) \!-\!f_t(\bm x_t) \Big)\!\right] \!\stackrel{(\alpha)}\leq \sumT\! \Big(\!G_f\E\left[|\bm x_t\! -\!\bm{x}_t^m| \right]\!\Big)\\
 &\stackrel{(\beta)}\leq G_f\sumT  \sqrt{\E\left[ \sum_{i=1}^I\sum_{j=1}^J\Big({x}_{ij}(t) -\E[{x}_{ij}(t)] \Big)^2\right]}.
\end{align*}
In $(\alpha)$ we used that expectation is a linear operator and the Lipschitz constant of $f_t$, and in $(\beta)$ Jensen's inequality and the unbiased sampling. This is the variance of the random binary output of $Q_{\mathcal X}$. Since the binary vector is subject to a simplex per user, each sum w.r.t. $j$ is bounded by the variance $1\!-\!(1/J)$, and the overall term by $\mu\!\triangleq\! \sqrt{I\!-\!(I/J)}$. Using \eqref{eq:regret-aux1}:
\begin{align*}
	\E[\mathcal R_T]&= G_fT\mu+\frac{1}{\beta}\ln{\frac{1}{w_1^k}}+\beta T\nu\notag \\
	&+(G^2+2G_A)^{1/2}(D_A^2+2D_{A*}P_T)^{1/2}.
\end{align*}
The last term can be made sublinear using the step normalization trick of \cite[Lem. 3]{lesage-tac-21}. Due to lack of space we omit the details of the derivation.

\bibliography{references-ho}
\bibliographystyle{IEEEtran}
\end{document}